\documentclass[aps,floats,twocolumn,showpacs]{revtex4-1}
\usepackage[cp1251]{inputenc}
\usepackage[english]{babel}
\usepackage{amssymb}
\usepackage{amsmath}
\usepackage{graphicx}
 \newcommand{\rot}{\mbox{\rm{rot\,}}}

 \def\bc{\begin{center}}          \def\ec{\end{center}}
 \allowdisplaybreaks

\begin{document}
 \title{Note on quantitatively correct simulations of the kinetic beam-plasma instability}
 \author{K.V.Lotov, I.V.Timofeev}
 \affiliation{Budker Institute of Nuclear Physics SB RAS, 630090, Novosibirsk, Russia}
 \affiliation{Novosibirsk State University, 630090, Novosibirsk, Russia}
 \author{E.A.Mesyats, A.V.Snytnikov, V.A.Vshivkov}
 \affiliation{Institute of Computational Mathematics and Mathematical Geophysics SB RAS, 630090, Novosibirsk, Russia}
 \date{\today}
 \begin{abstract}
A large number of model particles is shown necessary for quantitatively correct simulations of the kinetic beam-plasma instability with the clouds-in-cells method. The required number of particles scales inversely with the expected growth rate, as in the kinetic regime only a narrow interval of beam velocities is resonant with the wave.
 \end{abstract}
 \pacs{52.65.Rr, 52.35.Qz}
 \maketitle


Beam-plasma interaction plays an important role in various physical phenomena
such as transport of relativistic electrons in the fast ignition scheme of inertial fusion, gamma-bursts, solar type II and III radio bursts, and collisionless shock waves in the space plasma (see review \cite{bret} and references therein). Also, beam-plasma collective interaction determines the efficiency of turbulent plasma heating \cite{tim,bur} and electromagnetic emission \cite{thum,pos,tim1,arzh} in fusion-oriented mirror traps.

One of the most popular and effective tools for theoretical studies of the beam-plasma interaction is numerical simulations by the Particle-In-Cell (PIC) method. At present there is a great number of one-, two-, and three-dimensional PIC codes developed. These codes are used to reproduce fine details of the complex chain of intermediary processes that lead to plasma heating or electromagnetic radiation. The beam densities of interest $n_b$ are usually much smaller than the plasma density $n_p$. For the fast ignition problem, the electron beams are relatively dense ($n_b/n_p \sim 0.1$) \cite{grem,pukh,kong}. Weakly relativistic beams with $n_b/n_p \sim 10^{-4} \div 10^{-3}$ are interesting for mirror traps \cite{lot,tim2,tim3}. Non-relativistic beams of very low density, $n_b/n_p \sim 10^{-8} \div 10^{-5}$ are of interest for radiation generation in solar radio bursts \cite{tsi}, though simulated with higher beam densities because of code limitations \cite{rhee,yi,sau,tsi}. The numerical study usually begins from the linear stage of the beam-plasma instability, so the quantitatively correct simulation of this stage at low beam densities is important for the whole process.

There are many realizations of the PIC method that differ in the algorithm of charge and current evaluation from positions of model particles. This algorithm is usually referred to as the shape function. One of the simplest shape functions is the triangular one also called Clouds-In-Cells  (CIC) or linear interpolation \cite{CIC-first}. This shape function is still widely used \cite{tsi,CIC2,CIC3,CIC4,CIC5,AIP1507-416,PoP14-043103,PRST-AB10-034203} in spite of availability of more accurate algorithms \cite{Hockney-Eastwood,Birdsall+}.

In this paper we show that an uncommonly large number of model particles is necessary for quantitatively correct simulations of the kinetic beam-plasma two-stream instability with the CIC method. The smaller the growth rate the more particles are needed. If the number of particles is insufficient, then the results are only qualitatively correct, that is the beam-plasma system behaves realistically, but the growth rate is underestimated. To prove this statement, we simulate various regimes of the two-stream instability with a rather typical three-dimensional CIC code and compare the simulated growth rates with analytically obtained values.


We solve Vlasov equations for ions and electrons and Maxwell equations for the fields by the method described in Ref.~\cite{Vshivkov}. The units of measure are: the electron mass $m_e$ for masses, the light velocity $c$ for velocities, the elementary charge $e$ for charges, the unperturbed plasma density $n_0$ for densities, the inverse electron plasma frequency $\omega_p^{-1} = (4 \pi n_0 e^2/m_e)^{-1/2}$ for times, and the wavebreaking field $E_0 = m_e c \omega_p/e$ for fields. This normalization is commonly used in plasma wakefield acceleration and relativistic beam studies. The Vlasov equations are solved with the PIC method, that is we follow trajectories of macro-particles in the six-dimensional phase space with the leapfrog scheme:
\begin{multline}\label{ep1}
 \frac{\textbf{p}^{m+1/2} - \textbf{p}^{m-1/2}}{\tau} = \\
 q \left(\textbf{E}^m + \left[\frac{\textbf{v}^{m+1/2} + \textbf{v}^{m-1/2}}{2} ,\textbf{B}^m \right] \right),
\end{multline}
\begin{equation}\label{ep2}
 \frac{\textbf{r}^{m+1} - \textbf{r}^{m}}{\tau} = \textbf{v}^{m+1/2}, \phantom{mmmmmmmmmmmmm}
\end{equation}
where $\tau$ is the time step, the superscripts denote the time slices, and other notation is common. The fields are obtained with Langdon-Lasinski scheme \cite{Yee}:
\begin{gather}
\label{eqBE1}
\frac{{\textbf{B}}^{m+1/2}-{\textbf{B}}^{m-1/2}}{\tau} =- \rot_h \textbf{E}^m, \\
\label{eqBE2}
\frac{{\textbf{E}}^{m+1}-{\textbf{E}}^m}{\tau}= \rot_h \textbf{B}^{m+1/2} - \textbf{j}^{m+1/2}.
\end{gather}
Here $\rot_h$ is the curl operator defined in the grid space \cite[p.\,169]{Vshivkov}. The vectors of electric and magnetic fields are computed on shifted grids:
\begin{eqnarray}\label{eq:BE}
& & {\bf{B}}^{m+1/2}=(B^x_{i+1/2,j,l}, B^y_{i,j+1/2,l}, B^z_{i,j,l+1/2})^{m+1/2},\nonumber
\\
& & {\bf{E}}^{m}=(E^x_{i,j+1/2,l+1/2}, E^y_{i+1/2,j,l+1/2}, E^z_{i+1/2,j+1/2,l})^m,\nonumber
\\
& & {\bf{j}}^{m+1/2}=(j^x_{i,j+1/2,l+1/2},j^y_{i+1/2,j,l+1/2},j^z_{i+1/2,j+1/2,l})^{m+1},\nonumber
\\
& & \rho^m=\rho^m_{i+1/2,j+1/2,l+1/2}. \nonumber
\end{eqnarray}
Here the subscripts denote the position on the spatial grid, and superscripts $x,y,z$ are components of vectors. The scheme gives the second order of approximation in space and time.


As we aim at precise comparison of the growth rates, special care is taken to simulate the instability in its clearest form. The electron beam velocity distribution is
\begin{equation}\label{eq8}
    f(v_z) = \frac{1}{\sqrt{2\pi} \, \Delta v} \exp \left( -\frac{(v_z-v_0)^2}{2 \Delta v^2} \right)
\end{equation}
with $v_0=0.2$. The width of the simulation area is taken small to make the instability one-dimensional. The length $L_s$ of the simulation area is determined by the wavenumber $k_s$ of the mode of interest, $L_s = 2 \pi / k_s$, so that the periodic boundary conditions select only one unstable mode out of the continuous spectrum typical for the infinite plasma. To avoid generation of the magnetic field by the uncompensated beam current, we initiate model particles in quadruples. Each quadruple consists of one beam electron initiated with a purely longitudinal velocity $v_z$ in accordance with \eqref{eq8}, two plasma electrons with equal longitudinal velocities exactly compensating the beam current and opposite transverse velocities, and a plasma ion that exactly compensates the charge of the first three. The quadruples have the uniform random distribution over the simulation area. The required density ratio is ensured by weighting model particles. The transverse temperature of plasma electrons of the order of 0.5\,eV is chosen by try and error to suppress beam filamentation and have no visible effect on the longitudinal two-stream instability. The ion mass is 1836 (hydrogen). The spatial grid is $4 \times 4 \times 100$ nodes, the time step $\tau=0.001$. The grid size depends on the length of simulation window and is close to $0.01$. The number of beam model particles in the cell $l_p$ is the variable parameter dependence on which is under study.

For benchmarking, we use the growth rates obtained by exact numerical solution \cite{Vestnik3-62} of the beam-plasma dispersion relation in the one-dimensional case for several beam densities \cite{lot}. These known values $\gamma$ are then used for accuracy evaluation of the growth rates $\gamma_{1,2}$ obtained from PIC simulations.

We tested two methods of growth rate calculation on the basis of simulation output. Inherently, both of them are methods of early identification of the unstable wave. The first method is based on the exponential growth of the unstable wave energy $W \propto
e^{2\gamma t}$. The formula
\begin{equation}\label{eq:gamma1_0}
\gamma=\frac{1}{2}\frac{\partial }{\partial t} \ln W
\end{equation}
applied to simulation results becomes
\begin{equation}\label{eq:gamma1}
\gamma_1=\frac{1}{2}\frac{\partial }{\partial t} \ln \left[\sum\limits_{i,j,l=1}^{n_x \, n_y \, n_z} {\bf{E}} ^2(i,j,l,t) \right],
\end{equation}
where $n_s$ is the number of grid nodes in the direction $s$ $(s=x,y,z)$, and we used the obvious fact that the electric field energy grows with the same exponent as the total energy of the wave. Formula~(\ref{eq:gamma1}) turned out to be not the optimum one, as the whole spectrum of noise fields contributes the field energy and hampers identification of small growth rates.

The second method relies on the growth of a particular Fourier harmonic of the longitudinal electric field:
\begin{multline}\label{eq:gamma2}
\gamma_2=\frac{1}{2}\frac{\partial }{\partial t} \ln \left[ \left( \sum\limits_{i,j,l=1}^{n_x \, n_y \, n_z} E_z (i,j,l,t)\cos\left( \frac{2\pi l}{n_z} \right) \right)^2 \right. \\
+ \left. \left(\sum\limits_{i,j,l=1}^{n_x \, n_y \, n_z} E_z (i,j,l,t)\sin \left( \frac{2\pi l}{n_z} \right) \right)^2  \right].
\end{multline}
This method is more noise resistant, as all spatial harmonics of the noise field are ignored except one.

To demonstrate main findings, we take three test cases (Table~\ref{t1}) which correspond to hydrodynamic (1), transition (2), and kinetic (3) regimes of the two-stream instability. These regimes differ in beam fraction affected by the instability. In the hydrodynamic regime, the whole beam couples to the unstable wave. In the kinetic regime, only a small fraction of beam particles participates in wave excitation which is contained in a narrow velocity interval.
\begin{table}[htb]
 \caption{ Parameters of test cases.}\label{t1}
 \begin{tabular}{llll}\hline
    Case & 1 & 2 & 3 \\ \hline
    Beam density, $n_b$ & 0.002 & 0.002 & 0.0002 \\
    Beam velocity spread, $\Delta v/v_0$ & 0.035 & 0.14 & 0.14 \\
    Length of simulation area, $L_s$ \quad & 1.2566 \quad & 1.1424 \quad & 1.1424 \\
    Expected growth rate, $\gamma$ & 0.0706 & 0.0221 & 0.00256 \\
  \hline
 \end{tabular}
\end{table}

\begin{figure*}[p]
\includegraphics[width=300bp]{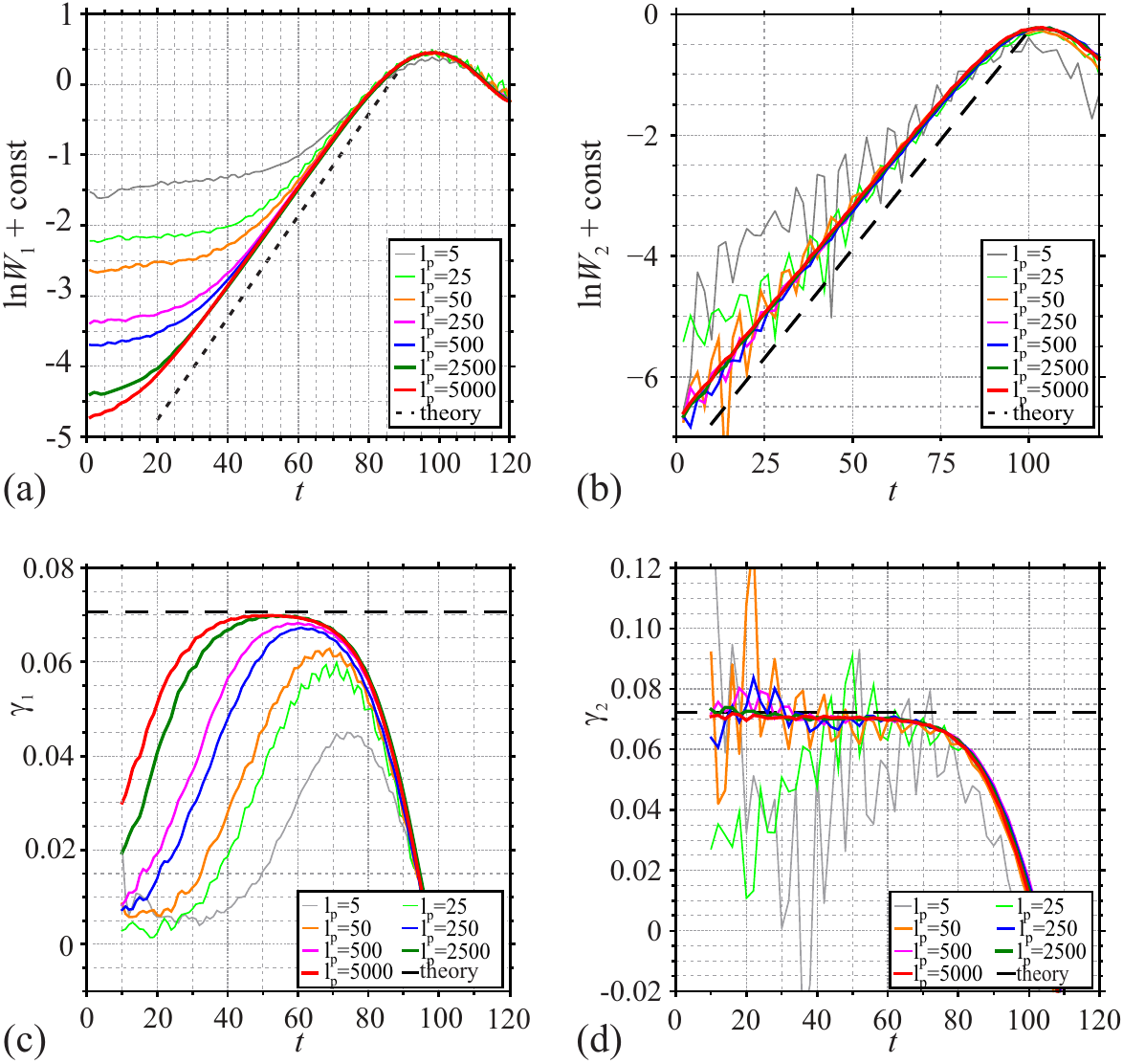}
\caption{(Color online) The total field energy $W_1$ (a), the energy of the unstable harmonic $W_2$ (b), and the corresponding growth rates (c,d) for the hydrodynamic instability regime simulated with different numbers of beam macro-particles. Dashed straight lines correspond to the expected growth rate; in fragments (a,b) these lines are shifted horizontally for better visibility.}\label{f1-hydro}
\end{figure*}

The hydrodynamic regime is the easiest for simulations, so we use it to compare two methods of growth rate evaluation. Shown in Fig.\,\ref{f1-hydro} are the quantities contained in square brackets in \eqref{eq:gamma1}, \eqref{eq:gamma2}; we denote them $W_1$ and $W_2$, respectively. The total energy $W_1$ has much shorter time interval of the exponential growth because of the higher level of the noise energy. Even though the plasma is initially cold, it is rapidly heated to the some equilibrium level \cite{Hockney-Eastwood} which depends on computational parameters. As a consequence, it is much more difficult to identify the unstable wave with the first method and find its growth rate. Therefore we use the second method of wave identification in what follows.
\begin{figure*}[p]
\includegraphics[width=300bp]{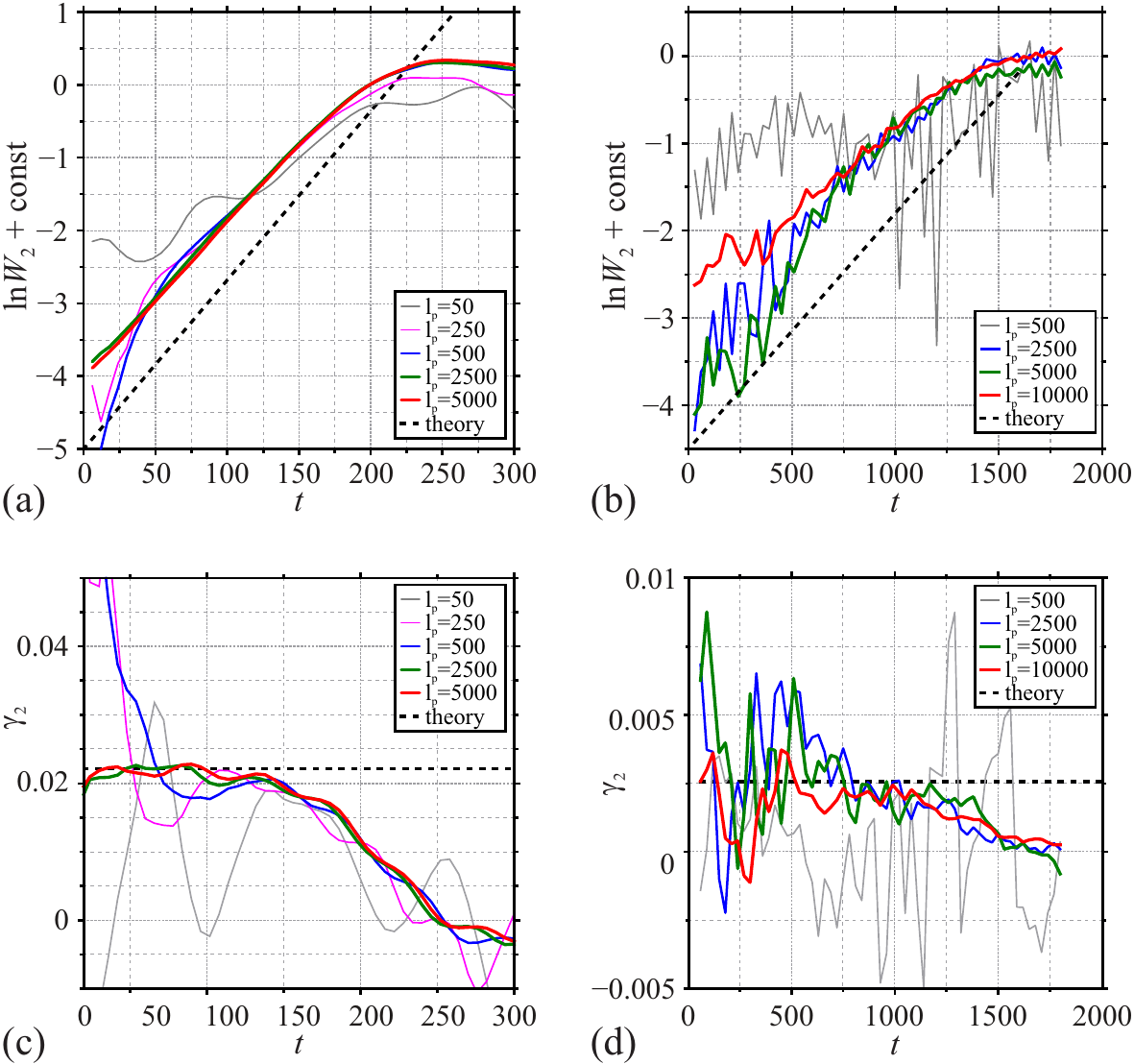}
\caption{(Color online) The energy of the unstable harmonic $W_2$ (a,b), and the corresponding growth rates (c,d) for the transition (a,c) and kinetic (b,d) instability regimes simulated with different numbers of beam macro-particles. Dashed straight lines correspond to the expected growth rate; in fragments (a,b) these lines are shifted horizontally for better visibility.}\label{f2-kinetic}
\end{figure*}

\begin{figure}[tb]
\includegraphics[width=150bp]{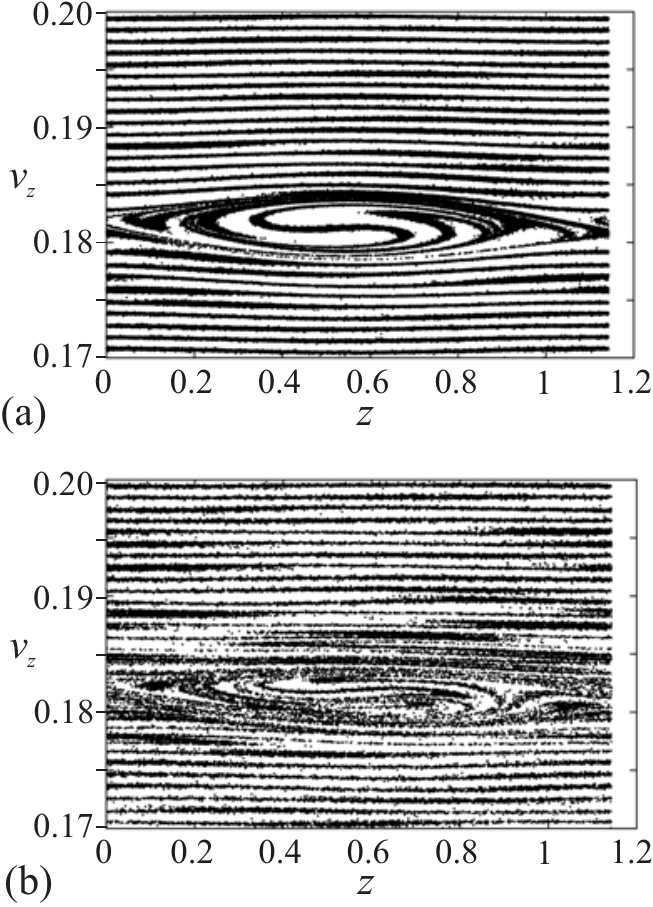}
\caption{Fragments of beam phase space portraits for sufficient with $l_p = 2500$ (a) and insufficient with $l_p = 500$ (b) number of macro-particles in the kinetic regime at $t=2100$.}\label{f3-phase}
\end{figure}

In transition and kinetic regimes, the exponential growth with the correct growth rate is observed for large number of particles in cell only (Fig.\,\ref{f2-kinetic}): $l_p \gtrsim 250$ for the regime 2 and $l_p \gtrsim 2500$ for the regime 3. To check whether the wave is simulated incorrectly at small $l_p$, we plot phase portraits of the beams perturbed by the grown wave in Fig.\,\ref{f3-phase}. These graphs are produced by drawing only those beam particles which initial velocity falls within selected equidistant intervals. At $t=0$ these particles form equidistant horizontal lines on the phase plane. As the instability develops, these lines curl up into the well known eye-like pattern from which the separatrix is clearly seen. If the number of model particles is insufficient, then the pattern is distorted. This means the incorrect growth rate comes from inaccurate simulations rather than from imperfect identification method.

Note that the necessary total number of model particles participating in formation of the eye-like structure is the same in regimes 2 and 3, about $2 \cdot 10^5$. Simulation series conducted with different grid sizes and simulation areas of different widths indicate that it is the total number of particles in the resonance area of the wave which determines whether the growth rate is correct or not. By the resonance area we mean one wave period in length, the whole wave packet in transverse dimensions, and the velocity interval
\begin{equation}\label{eq9}
    \delta v_z \approx 14 \gamma/k_s,
\end{equation}
which is the theoretically predicted \cite{lot} width of the separatrix at the highest (nonlinearly saturated) wave amplitude. Thus we can formulate the necessary condition for quantitatively correct three-dimensional simulations of the kinetic two-stream instability by the CIC method. The number of beam model particles resonant with the wave must be $\sim 10^5$ or higher.

Let us re-formulate the drawn conclusion in terms of particles in cell. The considered resolution (100 grid points per wave period) is rather typical for PIC simulations. Assuming the wave packet width is $\sim k_s^{-1}$ in both transverse directions (or 16 grid points), we find that $2 \cdot 10^5/100/16^2\sim 10$ particles in cell are sufficient, if the whole beam interacts with the wave in a wide simulation area. If, however, the instability is in the kinetic regime, and only a small fraction $\varepsilon$ of beam particles is expected to interact with a wave packet, then the required number of particles in cell is $\sim 10/\varepsilon$. We deliberately leave the main message of the paper in this loose form, since the exact criterion may differ for different problems and different realizations of the PIC method.

This work is supported by The Ministry of education and science of Russia (grant RFMEFI61914X0003) and by RFBR (grants 14-02-00294 and 14-07-00241).

\end{document}